\documentstyle[aps,pre,epsfig]{revtex}

\newcommand{\be}{\begin{equation}}
\newcommand{\ee}{\end{equation}}
\newcommand{\ba}{\begin{eqnarray}}
\newcommand{\ea}{\end{eqnarray}}

\title{\bf Measuring subdiffusion parameters}

\author{T. Koszto{\l}owicz$^1$, K. Dworecki$^1$, and 
St. Mr\'owczy\'nski$^{1,2}$}

\address{$^1$Institute of Physics, \'Swi\c etokrzyska Academy,
ul. \'Swi\c etokrzyska 15, PL - 25-406 Kielce, Poland \\
$^2$So\l tan Institute for Nuclear Studies,
ul. Ho\.za 69, PL - 00-681 Warsaw, Poland \\[2mm]}

\date{27-th January 2005}

\begin{document}
\maketitle

\begin{abstract}

We propose a method to extract from experimental data the subdiffusion 
parameter $\alpha$ and subdiffusion coefficient $D_{\alpha}$ which are 
defined by means of the relation $\langle x^2\rangle =\frac{2D_\alpha}
{\Gamma( 1+\alpha)}\,t^\alpha$ where $\langle x^2\rangle$ denotes a mean 
square displacement of a random walker starting from $x=0$ at the initial 
time $t=0$. The method exploits a membrane system where a substance of 
interest is transported in a solvent from one vessel to another across 
a thin membrane which plays here only an auxiliary role. Using such a system, 
we experimentally study a diffusion of glucose and sucrose in a gel solvent.
We find a fully analytic solution of the fractional subdiffusion equation
with the initial and boundary conditions representing the system under
study. Confronting the experimental data with the derived formulas, we show 
a subdiffusive character of the sugar transport in gel solvent. We precisely
determine the parameter $\alpha$, which is smaller than 1, and the subdiffusion 
coefficient $D_{\alpha}$.

\end{abstract}

 
\section{Introduction}
\label{Intro}


Subdiffusion occurs in various systems. As examples, we mention here  
a diffusion in porous media or charge carriers transport in amorphous 
semiconductors \cite{mk,mksb,bg}. The subdiffusion is characterized by 
a time dependence of the mean square displacement of a Brownian particle. 
When the particle starts form $x=0$ at the initial time $t=0$ this 
dependence in a one-dimension system is of the form 
\be
\label{a}
\langle x^{2}\rangle =\frac{2D_{\alpha}}
{\Gamma\left( 1+\alpha\right)} \: t^{\alpha}\;,
\ee
where $D_{\alpha}$ is the subdiffusion coefficient measured in the units 
${\rm [m^2/s^\alpha ]}$; the parameter $\alpha$, which we call here as 
a subdiffusion parameter, obeys $0< \alpha \le 1$. For $\alpha =1$ one 
deals with the normal or Gaussian diffusion. The linear growth of 
$\langle x^{2}\rangle$ with $t$, which is characteristic for normal 
diffusion, results from the Central Limit Theorem applied to many independent
jumps of a random walker. The anomalous diffusion occurs when the famous
theorem fails to describe the system because the distributions of summed 
random variables are too broad or the variables are correlated to each other. 
In physical terms, the subdiffusion is related to infinitely long average
time that a random walker waits to make a finite jump. Then, its average
displacement squared, which is observed in a finite time interval, is 
dramatically suppressed.

The subdiffusion has been recently extensively studied, see {\it e.g.} 
\cite{mk,mksb,bg,bmk,eb,lm,km}. While the phenomenon is theoretically 
rather well understood there are very a few experimental investigations. 
There is no effective method to experimentally measure the parameters 
$\alpha$ and $D_{\alpha}$. In the pioneering study \cite{km}, where the 
subdiffusion coefficient was determined experimentally for the first time, 
the interdiffusion of heavy and light water in a porous medium was observed 
by means of nuclear magnetic resonance. The subdiffusion coefficient 
$D_\alpha$ was determined, using the special case $\alpha = 2/3$ solution 
of the fractional derivative diffusion equation. The procedure is neither 
very accurate nor of general use.    

Our aim here is to develop a method to precisely measure the parameters 
$\alpha$ and $D_{\alpha}$. For practical reasons, which are explained below, 
we choose for the experimental study a membrane system containing two vessels
with a thin membrane in between which separates the initially homogeneous 
solute of the substance of interest from the pure solvent. A schematic
view of the system is presented in Fig.~1. We show that the membrane does 
not affect, as expected, the values of investigated parameters. 
Instead of the mean square displacement (\ref{a}), our method refers 
to the temporal evolution of the thickness $\delta$ of the so-called 
near-membrane layer which is defined as a distance from the membrane 
where the substance concentration drops $\kappa$ times with respect to the 
membrane surface; $\kappa$ is an arbitrary number. In our previous paper 
\cite{dkwm}, we demonstrated that for the normal diffusion 
$\delta(t) = A \sqrt{t}$. Here we are going to study the diffusion of 
glucose and sucrose in a gel solvent to show that 
\be
\label{aa}
\delta\left( t\right) = A \: t^{\gamma}\;,
\ee
with $\gamma <1/2$. Our choice of the gel medium to look for a subdiffusion 
is far not accidental. A gel is built of large and heavy molecules which 
form a polymer network. Thus, the gel water solvent resembles a porous 
material filled with water. Since a mobility of sugar molecules is highly 
limited in such a medium the subdiffusion is expected. As we show below, 
this expectation is indeed fulfilled.

Using an analytic solution of the fractional subdiffusion equation,
we theoretically argue, that $\alpha = 2\gamma$ and that the parameter $A$ 
uniquely depends, as in the case of the normal diffusion, on the number 
$\kappa$ and the parameters $\alpha$ and $D_{\alpha}$. Therefore, knowing $A$, 
$\kappa$ and $\alpha$, one can deduce $D_{\alpha}$ for arbitrary $\alpha$. 
Since the subdiffusion parameter $\alpha$ is measured with a high accuracy 
we can distinguish a very slow normal diffusion from a subdiffusion with 
$\alpha$ close to unity. Our very first theoretical and experimental results 
on the subdiffusion parameter $\alpha$ have been published in two short 
notes \cite{kd1}.


\section{Experiment}
\label{Exp}


The membrane system under investigation is the cuvette of two vessels
separated by the horizontally located membrane. Initially, we fill the 
upper (lower) vessel with the solute of transported substance while in 
the lower (upper) one there is a pure solvent. Then, the substance
diffuses from one vessel to another through the membrane. Since the 
concentration gradient is in the vertical direction only, the diffusion 
is expected to be one-dimensional (along the axis $x$). 

The substance concentration is measured by means of the laser interferometric 
method. The laser light is split into two beams. The first one goes through 
the membrane system parallelly to the membrane surface while the second 
(reference one) goes directly to the light detecting system. The interferograms,
which appear due to the interference of the two beams, are controlled by the 
refraction coefficient of the solute which is turn depends on the substance 
concentration. The analysis of the interferograms allows one to reconstruct 
the time-dependent concentration profiles of the substance transported in 
the system and to find the time evolution of the near-membrane layers 
which are of our main interest here. We note that the measurement does not 
disturb the system under study. The experimental set-up is described in 
detail in \cite{kd}. Here we only mention that it consists of the cuvette 
with membrane, the Mach-Zehnder interferometer including the He-Ne laser, 
TV-CCD camera, and the computerized data acquisition system.

For each measurement, we prepared two gel samples: the pure gel - 1.5\% 
water solution of agerose and the same gel dripped by the solute of glucose 
or sucrose. The concentration of both sugars in the gel was fixed to be
either $0.1 \; [{\rm mol/dm^3}]$ or $0.07 \; [{\rm mol/dm^3}]$. The two 
vessels of the membrane system were then filled with the samples and the 
(slow) processes of the sugar transport across the membrane started. We used
an artificial membrane of the thickness below 0.1 mm. The membrane was 
needed for two reasons. It initially separated the homogeneous sugar solute 
in one vessel from the pure gel in another one. It also precisely fixed the 
geometry of the whole system.
 
When the sugar was diffusing across the membrane we were recording the 
concentration profiles in the vessel which initially contained pure 
gel. The examples of typical interferograms and extracted concentration 
profiles are presented in the earlier paper \cite{kd}. The thickness 
of a near-membrane layer $\delta$ was calculated from the measured 
concentration profiles $C(x,t)$ according to the definition
\be
\label{nml}
C(\delta ,t) = \kappa \: C( 0^+,t)\;,
\ee
where $\kappa$ is an arbitrary number smaller than unity. In our analysis
we used $\kappa=$ 0.12, 0.08 and 0.05. Computing $\delta$ for various  $t$, 
we found the thickness of near-membrane layer as a function of time.  

In Fig.~2 we present $\delta$ as a function of time for the glucose and 
sucrose transported in a gel for the initial sugar concentration equal
$0.1 \; [{\rm mol/dm^3}]$. For the glucose we took three values of 
$\kappa=$ 0.12, 0.08 and 0.05 while for the sucrose $\kappa=0.08$. 
The errors of  $\delta(t)$  shown in the figure were found in the 
following way.  We performed 6 independent measurements of the
concentration profiles $C(x,t)$ at the same initial sugar concentration.
The thickness of the near membrane layer $\delta (t)$ was found for each
concentration profile. We note here that $\delta (t)$ is insensitive to 
a sizeable error of absolute normalization of $C(x,t)$ as it cancels
due to the definition (\ref{nml}). Having 6 values of $\delta$ for 
every $t$, we obtained the mean value of $\delta$ and the standard 
error. The final errors shown in Figs.~2 were obtained by further 
multiplying the standard errors by the Student-Fisher coefficient 
taken at a confidence level 95\% to include the effect of low statistics.

As seen in Fig.~2, the experimentally found time dependence of $\delta$ 
is well described by the power function (\ref{aa}) with the common index 
$\gamma = 0.45$. In Fig.~3 we show the same data as in Fig.~3 but in
a log-log scale to better test the power law dependence. The lines 
representing $\sqrt{t}$ are also shown for comparison. It is evident 
that the measured index $\gamma$ is smaller than 0.5. There are some 
deviations of our data from the power law dependence at $t < 300\:$s. 
However, as explained in Sec.~IIIB, our theoretical formulas, in particular 
the power law behavior, are derived in the long time approximation.

We fitted the experimental data shown in Figs.~2, 3 by the power law formula 
$A\, t^\gamma$, using a standard $\chi^2$ procedure. The values of $\chi^2$ 
per degree of freedom were smaller than unity for every value of $\kappa$. 
The error of the universal index $\gamma = 0.45$ was found to be 0.005.
The parameter $A$ depends on $\kappa$; it is also different for glucose 
than for sucrose. We found
\begin{eqnarray}
\label{A-g1}
A &=& 0.091 \pm 0.004 \;\;\;\;\;{\rm for}\;\;\;\;\;\kappa=0.05 \;, \\
\label{A-g2}
A &=& 0.081 \pm 0.004 \;\;\;\;\;{\rm for}\;\;\;\;\;\kappa=0.08 \;, \\
\label{A-g3}
A &=& 0.071 \pm 0.004 \;\;\;\;\;{\rm for}\;\;\;\;\;\kappa=0.12 \;,
\end{eqnarray}
for glucose and
\begin{eqnarray}
\label{A-s1}
A = 0.064 \pm 0.003 \;\;\;\;\;{\rm for}\;\;\;\;\;\kappa=0.08 \;.
\end{eqnarray}
for sucrose.

The fact that $\gamma$ is the same for glucose and sucrose suggests
that the index depends only on the solvent {\it i.e.} the medium where
the subdiffusion occurs. As will be shown in Sec.~\ref{Exp-D-a}, the
subdiffusion coefficients for glucose and for sucrose differ from each
other. Such a situation resembles the case of normal diffusion where
$\gamma=1$ is universal but the diffusion coefficient changes from one
substance to another.

At the end of this section we note that the results of measurements 
of the near-membrane layers for the initial sugar concentration equal
$0.07 \; [{\rm mol/dm^3}]$ fully coincide with those presented above
which were performed for the initial concentration equal 
$0.1 \; [{\rm mol/dm^3}]$. This simply reflects the linearity of 
the subdiffusion equation which is discussed in the next section.


\section{Theory}
\label{Theo}


In this section we discuss a theoretical description of the experimental
result presented in Sec.~\ref{Exp}. In particular, we derive the formula
which allows one to obtain the subdiffusion coefficient $D_\alpha$
from the experimental data.

\subsection{General formulas}

The subdiffusion is described by the equation  
\cite{mk,compte}
\be
\label{se}
\frac{\partial C( x,t)}{\partial t} =
D_{\alpha}\frac{\partial^{1-\alpha}}{\partial t^{1-\alpha}}
\frac{\partial^{2}C\left( x,t\right)}{\partial x^{2}}\;,
\ee
with the Riemann-Liouville fractional derivative defined as
$$
\frac{\partial^\alpha f(t)}{\partial t^\alpha} =
\frac{1}{\Gamma (n - \alpha )} 
\frac{\partial^n }{\partial t^n}
\int_0^t dt' \, \frac{f(t')}{(t - t')^{1 + \alpha -n}} \;,
$$
where $n$ is the smallest integer larger than $\alpha > 0$.
It is tempting to write down Eq.~(\ref{se}) in a simpler form as
\be 
\label{se1}
\frac{\partial^\alpha C(x,t)}{\partial t^\alpha} =
D_{\alpha}\,
\frac{\partial^{2}C(x,t)}{\partial x^{2}}\;.
\ee
In general, however, the fractional derivative does not obey the
property
$$
\frac{\partial^\alpha }{\partial t^\alpha}
\frac{\partial^\beta }{\partial t^\beta}
= \frac{\partial^{(\alpha + \beta)} }{\partial t^{(\alpha + \beta)}} \;,
$$
and Eq.~(\ref{se}) is not fully equivalent to Eq.~(\ref{se1}) \cite{Old74}. 
And because the subdiffusion equation as derived in the framework of 
Continuous Time Random Walk \cite{compte} is of the form (\ref{se}), 
it has become customary to use  Eq.~(\ref{se}) rather than 
Eq.~(\ref{se1}). We also note here that Eq.~(\ref{se}) with 
the temporal fractional derivative of order $\alpha < 1$ corresponds 
to a `long' (infinite on average) waiting time of the random walker. 
This is just the physical situation which is expected in a gel 
medium built of large and heavy molecules forming a polymer network 
where mobility of the walker is strongly limited.

We solve Eq.~(\ref{se}) in the region $x>0$ for the following initial 
concentration 
\ba
\label{ic}
C( x,0) =\bigg\{
\begin{array}{ccr}
C_{0} & , & x<0\;,\\
0 & , & x>0\;.
\end{array}
\ea
Here, $x=0$ is the position of an infinitely thin membrane. In fact, we 
solve Eq.~(\ref{se}) for the Green's function $G(x,t;x_0)$ with the initial 
condition
\be
\label{icgf}
G(x,t=0;x_0)=\delta(x-x_0)\;,
\ee
and then, the concentration profiles are calculated using the integral formula
\be
\label{int}
C(x,t) =\int  dx_0 \: G(x,t;x_0) \: C(x_0,0) \;.
\ee
The Green's function $G(x,t;x_0)$ gives the probability density to find 
a random walker at the position $x$ in time $t$; the walker starts from 
$x_0$ at $t=0$.

To find the concentration profile $C(x,t)$ and then time evolution of 
near-membrane layer $\delta$, we use the relation  
\be
\label{gint}
G(x,t;x_0) =\int_0^{t} dt'\:  J(0^+,t';x_0) \: 
G_{\rm ref}( x,t-t';0^+) \;,
\ee
where $x > 0$ while $x_0 < 0$; $J(x,t;x_0)$ is the flux associated 
with $G(x,t;x_0)$ which for $x=0$  gives the flow across the membrane; 
$G_{\rm ref}(x,t;x_0)$ is the Green's function for the half-space system 
with $x > 0$ and the fully reflecting wall at $x=0$. The integral formula 
(\ref{gint}), which we call the wall relation, represents one-half of the 
membrane system as a half-space system with a reflecting wall replacing 
the membrane. The substance flux, however, does not vanish at the wall but 
it equals the actual flux in the membrane system. For the case of normal 
diffusion the wall relation (\ref{gint}) was used in Ref.~\cite{OSK}. 
In Appendix \ref{appendix1} we show that it also holds for subdiffusion.    

Using the relation (\ref{gint}) and Eq.~(\ref{int}), the concentration
profile can be written as
\be
\label{prof}
C(x,t) =\int_0^t dt' \, W(t') \:
G_{\rm ref}(x,t-t';0^+) \;,
\ee
where $x > 0$ and the function $W(t)$, which equals
\be 
\label{W-def}
W(t) =\int_{- \infty}^0 dx_0 \,  J(0^+,t;x_0) \:
C( x_0,0) \;,
\ee
contains all information about the initial and boundary conditions.
The interval of integration in Eq.~(\ref{W-def}) is due to the initial
condition (\ref{ic}).

Since the subdiffusion equation is of the second order with respect to 
the space variable $x$, it requires two boundary conditions at the membrane. 
The first one simply assumes the continuity of the flux $J$ flowing through 
the membrane {\it i.e.}
\be
\label{con}
J(0^-,t) =J(0^+,t)\;,
\ee
where $J$ is the subdiffusion flux given by the generalized Fick's law
\cite{z}.  It gets a simple form after Laplace transformation 
$L\left\{ f( t)\right\} \equiv \hat{f}( s) =\int_{0}^{\infty}f( t)e^{-st}$. 
Then,
it reads \cite{z}
\be
\label{current}
\hat{J}(x,s)=-D_{\alpha}\: s^{1-\alpha} \: \frac{d\hat{C}(x,s)}{dx}\;.
\ee

There is no obvious choice of the second boundary condition. Therefore,
we assume that the missing condition is given by a linear combination of
concentrations and flux at the membrane {\it i.e.}
\be
\label{general-boundary}
b_1 C(0^-,t) + b_2 C(0^+,t) + b_3 J(0,t) = 0 \;.
\ee
Since the current is continuous at the membrane there is no point to 
distinguish $J(0^+,t)$ from $J(0^-,t)$. We note that two boundary conditions
discussed in the literature are of the general form (\ref{general-boundary}).  
The first condition can be formulated as follows: {\em If during a given 
time interval, $N$ particles reach the membrane, the fraction $\sigma$ of 
them will be stopped while $( 1-\sigma)$ will go through.} This condition 
leads to the boundary condition of the form \cite{dkwm,kpr,kjp,kp1}
\be
\label{nbc}
C(0^+,t)=\frac{1-\sigma}{1+\sigma} \: C(0^{-},t)\,.
\ee
The second condition assumes that the flux flowing through the membrane
is proportional to the difference of concentration
at membrane surfaces  \cite{kmapp,kp}
\be
\label{rbc}
J(0,t) =\lambda \left( C(0^-,t) -C(0^+,t)\right)\,,
\ee
where the parameter $\lambda$ controls the membrane permeability. We call 
the boundary conditions (\ref{con},\ref{nbc}) and (\ref{con},\ref{rbc}) 
as `A' and `B', respectively, and in Sec.~\ref{spec-bc}, we use the indices 
`A' and `B' to denote the solutions of the fractional subdiffusion 
equation obeying the corresponding boundary conditions.

The initial condition (\ref{ic}) combined with the general boundary 
condition (\ref{general-boundary}) give the Laplace transform of the 
function $W(t)$, which is defined by Eq.~(\ref{W-def}), as
\be
\label{W}
\hat W (s) = C_0 \frac{b_1 \sqrt{D_\alpha}}
{b_1 - b_2 - b_3 \sqrt{D_\alpha} s^{1-\alpha/2}}
\, \frac{1}{s^{\alpha/2}}
= C_0 \frac{b_1 \sqrt{D_\alpha}}{b_1 - b_2}
\, \frac{1}{s^{\alpha/2}} 
\sum_{k=0}^{\infty}d^k s^{k(1-\alpha/2)} \;,
\ee
where 
$$
d \equiv \frac{b_3}{b_1 - b_2} \, \sqrt{D_\alpha} \;.
$$
The derivation of Eq.~(\ref{W}) is discussed in Appendix \ref{appendix3}. 
Inverting the Laplace transform, we get
\be
\label{W-exp}
W (t) = C_0 \frac{b_1 \sqrt{D_\alpha}}{b_1 - b_2}
\frac{1}{t^{1-\alpha/2}}
\sum_{k=0}^{\infty}\frac{d^k}{\Gamma(\alpha/2 - k(1-\alpha/2))} 
\frac{1}{t^{k(1-\alpha/2)}}\;.
\ee

The Green's function $G_{\rm ref}$, which enters Eq.~(\ref{prof}), 
can be easily obtained by means of the method of images \cite{mk} as 
\begin{equation}
\label{gref}
G_{\rm ref}(x,t;x_0) =G_0(x,t;x_0) +G_0(-x,t;x_0)\:,
\end{equation}
with $G_0$ being the Green's function for the homogeneous system given by
\begin{eqnarray}
\label{go}
G_0(x,t;x_0) =\frac{2\sqrt{D_{\alpha}}}{\alpha|x-x_0|} \:
H_{1\;1}^{1\;0}
\left(\left(\frac{|x-x_{0}|}{\sqrt{D_{\alpha}t^{\alpha}}}
\right)^{\frac{2}{\alpha}} \bigg| 
\begin{array}{cc}
1 & 1 \\
1 & \frac{2}{\alpha}
\end{array}
\right)\;,
\end{eqnarray}
where $H$ denotes the Fox function. In our numerical calculations we
use the form of $H$ 
\begin{equation}
\label{fh}
H_{1\;1}^{1\;0}\left( \frac{a^{\frac{1}{\beta}}}{t} \bigg| 
\begin{array}{cc}
1 & 1 \\
\frac{1+\nu}{\beta} & \frac{1}{\beta}
\end{array}
\right)=\beta\left(\frac{a^{\frac{1}{\beta}}}{t}\right)^{1+\nu}
\sum_{k=0}^{\infty}\frac{1}{\Gamma (-k\beta -\nu)k!}
\left (-\frac{a}{t^{\beta}}\right )^{k}\;,
\end{equation}
which is derived in \cite{koszt04} by means of the Mellin transform 
technique \cite{s}. In terms of the Laplace transform, the function 
(\ref{go}) simplifies to the form \cite{mk}
\begin{equation}
\label{golt}
\hat{G}_{0}(x,s;x_0) =\frac{1}{2\sqrt{D_{\alpha}}\, s^{1-\alpha/2}}
\: e^{-\mid x-x_{0}\mid \sqrt{\frac{s^{\alpha}}{D_{\alpha}}}}\;,
\end{equation}
which was found using the formula \cite{koszt04}
\ba
\label{inv-Lap1}
L^{-1}(s^\nu e^{-as^\beta}) =
\frac{1}{\beta a^{\frac{1+\nu}{\beta}}} \:
H_{1\;1}^{1\;0}\left( \frac{a^{\frac{1}{\beta}}}{t} \bigg|
\begin{array}{cc}
1 & 1 \\
\frac{1+\nu}{\beta} & \frac{1}{\beta}
\end{array}
\right) \;,
\ea
where $a,\beta > 0$. 

Having the explicit form of the functions $W$ and $G_{\rm ref}$,
we can write down the concentration profile using Eq.~(\ref{prof}).
The Laplace transform of $C$ equals
\be
\label{prof-s}
\hat C(x,s) = C_0 
\frac{b_1}{b_1 - b_2 - b_3 \sqrt{D_{\alpha}}\,s^{1-\alpha/2}}\,
\frac{1}{s}
\: e^{-x \sqrt{\frac{s^{\alpha}}{D_{\alpha}}}}\;,
\ee
which after inverting the transformation can be written down as
\be
\label{profil-final} 
C(x,t) = \int_0^t dt'\: W(t-t') \:
\frac{2}{\alpha x} \:
H_{1\;1}^{1\;0} \left(\left(\frac{x}{\sqrt{D_{\alpha}{t'}^{\alpha}}}
\right)^{\frac{2}{\alpha}} \bigg|
\begin{array}{cc}
1 & 1 \\
1 & \frac{2}{\alpha}
\end{array}
\right)\;.
\ee
Eq.~(\ref{profil-final}) is the starting point of our further analysis
which provides the function $\delta (t)$.


\subsection{Long time approximation}


We first consider the long time approximation of the formula 
(\ref{profil-final}) which, according to the Tauberian theorem
\cite{Sneddon72}, corresponds to the small $s$ limit of the Laplace 
transform.  The physical meaning of this approximation is discussed 
below. Taking into account only the leading contribution in the small 
$s$ limit, Eq.~(\ref{prof-s}) gets the form
\be
\label{prof-s-long}
\hat C(x,s) = C_0
\frac{b_1}{b_1 - b_2} \, \frac{1}{s}
\: e^{-x \sqrt{\frac{s^{\alpha}}{D_{\alpha}}}}\;,
\ee
which after inverting Laplace transformation provides
\be
\label{profil-long}
C(x,t) = C_0 \frac{2 b_1}{(b_1 -b_2)\alpha} \:
H_{1\;1}^{1\;0} \left(\left(\frac{x}{\sqrt{D_{\alpha}t^{\alpha}}}
\right)^{\frac{2}{\alpha}} \bigg|
\begin{array}{cc}
1 & 1 \\
0 & \frac{2}{\alpha}
\end{array}
\right)\;.
\ee
The solution (\ref{profil-long}) can be also obtained directly from 
Eq.~(\ref{profil-final}), taking into account only the $k=0$ term in 
the expansion of $W(t)$ given by Eq.~(\ref{W-exp}), and using the
formula
\be
\label{int-formula}
\frac{\sqrt{D_{\alpha}}}{x \Gamma(\alpha /2)} 
\int_0^t \frac{dt'}{(t-t')^{1-\alpha/2}} \:
H_{1\;1}^{1\;0} \left(\left(\frac{x}{\sqrt{D_{\alpha}{t'}^{\alpha}}}
\right)^{\frac{2}{\alpha}} \bigg|
\begin{array}{cc}
1 & 1 \\
1 & \frac{2}{\alpha}
\end{array} \right)
= H_{1\;1}^{1\;0} \left(\left(\frac{x}{\sqrt{D_{\alpha}t^{\alpha}}}
\right)^{\frac{2}{\alpha}} \bigg|
\begin{array}{cc}
1 & 1 \\
0 & \frac{2}{\alpha}
\end{array} \right) \;,
\ee
which is derived by means of the Laplace transformation.

The series (\ref{W-exp}) can be approximated by the first term if 
$d \ll t^{1-\alpha/2}$ which gives 
\be
\label{long-cond}
\bigg( \sqrt{D_\alpha} \: \frac{b_3}{b_1 - b_2} 
\bigg)^{\frac{1}{1-\alpha/2}} \ll t \;.
\ee 
When the boundary condition is of the form (\ref{nbc}), the condition
(\ref{long-cond}) is trivially satisfied for any $t$ as $b_3=0$ in this
case. Even more, as discussed in the next section,  the solution
(\ref{profil-long}) is {\em exact} for the boundary condition (\ref{nbc}).
When the boundary condition is of the form (\ref{rbc}), we have 
$\lambda = b_1/b_3 = - b_2/b_3$, and the condition (\ref{long-cond}) 
can be written as
\be
\label{long-cond2}
\bigg(\frac{\sqrt{D_\alpha}}{2 \lambda} 
\bigg)^{\frac{1}{1-\alpha/2}} \ll t \;.
\ee
We have first estimated the l.h.s. of Eq.~(\ref{long-cond2}),
assuming that we deal with the normal diffusion ($\alpha = 1$). 
For the membranes used in our experiments the parameter $\lambda$ is 
of order $10^{-2} \; {\rm [mm/s]}$ while the coefficient of normal 
diffusion $D$ is roughly $10^{-5} \; {\rm [mm^2/s]}$. Thus, the l.h.s. 
of Eq.~(\ref{long-cond2}) is estimated as 2~s. Since 10~s is the time 
step of our measurements which extend to 2500~s, the condition 
(\ref{long-cond2}) is fulfilled. We have also checked the condition 
(\ref{long-cond2}) {\it a posteriori}, using the values of $\alpha$ 
and $D_\alpha$ found in Sec.~\ref{Exp-D-a}. The l.h.s. of 
Eq.~(\ref{long-cond2}) is again about 2~s. 

Let us now discuss the temporal evolution of near-membrane layers in the 
long time approximation. Substituting the solution (\ref{profil-long}) 
into Eq.~(\ref{nml}), which defines the near-membrane layer, we get 
the equation
\be
\label{nml-eq0}
H_{1\;1}^{1\;0} \left(\left(\frac{\delta}{\sqrt{D_{\alpha}t^{\alpha}}}
\right)^{\frac{2}{\alpha}} \bigg|
\begin{array}{cc}
1 & 1 \\
0 & \frac{2}{\alpha}
\end{array} \right)
= \kappa 
H_{1\;1}^{1\;0}\left( 0 \Big|
\begin{array}{cc}
1 & 1 \\
0 & \frac{2}{\alpha}
\end{array}
\right)\;,
\ee
which due to the identity 
$$
H_{1\;1}^{1\;0}\left( 0 \Big|
\begin{array}{cc}
1 & 1 \\
0 & \frac{2}{\alpha}
\end{array}
\right)=\frac{\alpha}{2}\;,
$$
simplifies to 
\be 
\label{nml-eq}
H_{1\;1}^{1\;0} \left(\left(\frac{\delta}{\sqrt{D_{\alpha}t^{\alpha}}}
\right)^{\frac{2}{\alpha}} \bigg|
\begin{array}{cc}
1 & 1 \\
0 & \frac{2}{\alpha}
\end{array} \right)
= \frac{\kappa \alpha}{2}\;.
\ee

One observes that Eq.~(\ref{nml-eq}) is solved by
\be
\label{nml-sol}
\delta(t) = A(\alpha,D_{\alpha},\kappa) \: t^{\alpha/2} \;,
\ee
where the coefficient $A$ equals
\be
\label{an}
A(\alpha,D_{\alpha},\kappa) =
\sqrt{D_{\alpha}}\left[ \left(H_{1\;1}^{1\;0}\right)^{-1}
\left( \frac{\alpha\kappa}{2} \Big|
\begin{array}{cc}
1 & 1 \\
0 & \frac{2}{\alpha}
\end{array}
\right) \right]^{\frac{\alpha}{2}}\;.
\ee
We note that the near-membrane layer given by Eq.~(\ref{nml-sol})
does not depend on parameters $b_1$ and $b_2$ which control membrane 
permeability. $\delta (t)$ remains the same in the absence of membrane.
We also observe that the coefficient $A$ can be recalculated into the 
diffusion constant $D_\alpha$ as
\be
\label{sc}
D_{\alpha}=\frac{A^2}{\left[ \left(H_{1\;1}^{1\;0}\right)^{-1}
\left( \frac{\alpha\kappa}{2} \bigg|
\begin{array}{cc}
1 & 1 \\
0 & \frac{2}{\alpha}
\end{array}
\right) \right]^{\alpha}}\;.
\ee
Thus, knowing experimental values of $A$, $\alpha$ and $\kappa$, 
one can deduce $D_\alpha$.

\subsection{Specific boundary conditions}
\label{spec-bc}

In the previous section, we have determined the temporal evolution
of near-membrane layer for arbitrary linear boundary condition 
(\ref{general-boundary}) in the long time approximation. Here we discuss
the problem for two specific sets of the boundary conditions 
(\ref{con},\ref{nbc}) and (\ref{con},\ref{rbc}) which are called 
`A' and `B', respectively. The solutions of the subdiffusion equation 
(\ref{se}) satisfying the boundary conditions A and B can be obtained 
from the general solution (\ref{prof-s}), substituting 
$b_1 = \sigma + 1, \;\;  b_2 = \sigma -1, \;\; b_3 =0$ or 
$b_1 = \lambda, \;\; b_2 = -\lambda, \;\; b_3 = -1$, respectively. 
However, the solutions of Eq.~(\ref{se}) with the boundary conditions 
A and B can be also found in a much simpler as demonstrated in 
Appendix~\ref{appendix2}.

As in the case of normal diffusion, the boundary conditions 
(\ref{con},\ref{nbc}) allow for analytic solution of Eq.~(\ref{se}). 
Using the technique of Laplace transform (see Appendix~\ref{appendix2} 
for details), the concentration profile for $x>0$ is found as
\begin{equation}
\label{cpn}
C_{\rm A}(x,t) =(1-\sigma) \: \frac{C_{0}}{\alpha} \: 
H_{1\;1}^{1\;0} 
\left(\left(\frac{x^{2}}{D_{\alpha}t^{\alpha}}\right)^{\frac{1}{\alpha}}
\bigg| 
\begin{array}{cc}
1 & 1 \\
0 & \frac{2}{\alpha}
\end{array}
\right)\;.
\end{equation}
The solution (\ref{cpn}) plugged into the definition of near-membrane
layer (\ref{nml}) gives Eq.~(\ref{nml-eq}) which is solved by 
$\delta_{\rm A} (t)$ of the form (\ref{nml-sol}). Thus, as already 
mentioned, the formulas derived in the long time approximation are 
exact for the boundary condition A.

When Eqs.~(\ref{con}, \ref{rbc}) are used as the boundary conditions, the 
solution of subdiffusion equation (\ref{se}) can be written in the form 
of the infinite series of Fox functions. As explained in 
Appendix~\ref{appendix2}, for $x>0$ one finds
\begin{equation}
\label{cr}
C_{\rm B}(x,t) =\frac{C_0}{\alpha} \: 
\sum_{n=0}^{\infty}\left[ -\frac{x}{2\lambda}
\left( \frac{\sqrt{D_{\alpha}}}{x}\right)^\frac{2}{\alpha}\right]^n \:
H_{1\;1}^{1\;0}
\left(\left(\frac{x^{2}}{D_{\alpha}t^{\alpha}}\right)^{\frac{1}{\alpha}}
\bigg| 
\begin{array}{cc}
1 & 1 \\
n\left(\frac{2}{\alpha}-1\right) & \frac{2}{\alpha}
\end{array}
\right)\;.
\end{equation}

The solutions (\ref{cpn}) and (\ref{cr}), which for normal diffusion have
been discussed in \cite{kpr,kjp,kp1,kmapp}, qualitatively differ from each 
other. In particular, according to Eq.~(\ref{cpn}) the flux flowing 
through the membrane is constant in time while the solution (\ref{cr}) 
leads to the flow which decreases in time as the concentrations at both 
sides of the membrane evolve to the same value $C_0/2$. In the case of
Eq.~(\ref{cpn}), the ratio of concentrations at both sides is constant
in time as dictated by the boundary condition A. However, the qualitative
differences of the solutions (\ref{cpn}) and (\ref{cr}) are evident 
for times which are so long that substance concentration becomes 
approximately homogeneous at each side of the membrane in its 
neighborhood. These times are usually much longer than those satisfying 
the condition (\ref{long-cond}) when the vessels of a membrane system 
are of a few centimeters length as those used in our measurements. 

Since the formula (\ref{cr}) is analytically intractable we have found
the time evolution of near-membrane layer numerically. The results are
shown in Figs.~4-8. The physical meaning of the near-membrane layer 
suggests that it does not depend on the membrane permeability. The 
analytical calculations performed with the boundary condition A fully 
confirm this intuition. As seen, $\delta_{\rm A}$ remains the same even 
in the absence of the membrane ($\sigma = 0$). We are now going to show 
numerically that the same holds for the boundary condition B. Specifically, 
we argue that $\delta_{\rm A}(t) = \delta_{\rm B}(t)$ for the same values 
of $\kappa$, $\alpha$ and $D_\alpha$, in spite of the qualitative 
differences between the solutions (\ref{cpn}) and (\ref{cr}). 

Fig.~4 demonstrates that $\delta_{\rm B}(t)$ is independent of the 
membrane permeability parameter. Our numerical calculations, which are 
presented in Figs.~5-8, show that the thickness of the near-membrane 
layer grows in time as $t^\gamma$ with $\gamma = \alpha/2$. Thus, we write
\begin{equation}
\label{nmlr}
\delta_{\rm B}(t)=A_{\rm B} \: t^{\gamma}\;.
\end{equation}

We are now going to show that $A_{\rm A} = A_{\rm B}$.
In Figs.~5-7, we compare $\delta_{\rm B} (t)$ obtained numerically 
with $\delta_{\rm A} (t)$ given by Eq.~(\ref{nml-sol}) for the same values 
of $\kappa$, $\alpha$ and $D_{\alpha}$. In each of these figures one 
parameter changes whereas the remaining two are fixed. In the Fig.~5 we 
examine the dependence on the parameter defining thickness of near
membrane layer $\kappa$; in Fig.~6 there are several values of the 
subdiffusion coefficient $D_{\alpha}$ while in Fig.~7 the functions 
are plotted for several values of $\alpha$. In all these cases we see 
a perfect coincidence of $\delta_{\rm A} (t)$ with $\delta_{\rm B} (t)$. 
Fig.~8 demonstrates that $\delta_{\rm B} (t)$ fulfills the scaling exactly 
obeyed by $\delta_{\rm A} (t)$. Namely, we show that 
$\delta' (t)= \delta_{\rm B}(t) /A_{\rm A}$ with $A_{\rm A}$ given by 
Eq.~(\ref{an}) depends solely on time, exactly as 
$\delta_{\rm A} (t)/ A_{\rm A}$.

We conclude this section by saying that no difference between 
$\delta_{\rm A} (t)$ and $\delta_{\rm B} (t)$ has been observed.
Therefore, we expect that $A_{\rm A} = A_{\rm B} = A$. Consequently, 
the subdiffusion coefficients calculated from Eq.~(\ref{sc}) have to agree 
with the ones obtained numerically form the solution (\ref{cr}). Although 
we are unable to prove beyond the long time approximation that the time 
evolution of near-membrane layers does not depend on the boundary condition,
the results presented in this section strongly substantiate such a conjecture, 
and thus they justify the use of the formula (\ref{sc}) to evaluate the 
subdiffusive coefficient $D_{\alpha}$ from experimental data.


\section{Experimental values of $\alpha$ and $D_{\alpha}$}
\label{Exp-D-a}

In Sec.~\ref{Exp} we have fitted the experimentally obtained $\delta (t)$ 
by the power function $A\,t^\gamma$. Thus, we have found the index 
$\alpha = 2\gamma = 0.90 \pm 0.01$ and the values of the coefficient 
$A$ given in Eqs.~(\ref{A-g1},\ref{A-g2},\ref{A-g3}) and (\ref{A-s1}). 
Now, we recalculate $A$ into $D_\alpha$ by means of the relation (\ref{sc}). 
Using the numerical values of inverse Fox functions
\begin{eqnarray*}
\left( H_{1\;1}^{1\;0}\right)^{-1}
\left( 0.054 \Big| 
\begin{array}{cc}
1 & 1 \\
0 & \frac{1}{0.45}
\end{array}
\right) =6.032\;, \\ [2mm]
\left( H_{1\;1}^{1\;0}\right)^{-1}
\left( 0.036 \Big| 
\begin{array}{cc}
1 & 1 \\
0 & \frac{1}{0.45}
\end{array}
\right) =8.014\;, \\ [2mm]
\left( H_{1\;1}^{1\;0}\right)^{-1}
\left( 0.0225 \Big| 
\begin{array}{cc}
1 & 1 \\
0 & \frac{1}{0.45}
\end{array}
\right) =10.510\;, 
\end{eqnarray*}
which are calculated with the help of the expansion formula (\ref{fh}),
we computed the subdiffusion coefficient for each value of $\kappa$.
As expected, $D_\alpha$ shows no dependence on $\kappa$ within 
the errors which were obtained propagating the errors of $A$. 
We took the mean value of $D_\alpha$ as a final result but the 
final error is the maximal one. The point is that the values of 
$D_\alpha$ found for different values of $\kappa$ cannot be
treated as independent measurements. Thus, we obtained
$$
D_{0.90}=(9.8 \pm 1.0)\times 10^{-4}\; [{\rm mm^2/s^{0.90}}] 
$$
for glucose and
$$
D_{0.90}=(6.3 \pm 0.9)\times 10^{-4}\; [{\rm mm^2/s^{0.90}}] 
$$
for sucrose.

To be sure that Eq.~(\ref{an}), which is used to evaluate $D_\alpha$, 
properly describes the experimentally found $\delta (t)$, we checked 
the scaling property of $\delta (t)$ suggested by the formula 
(\ref{nml-sol}). In Fig.~9 we plotted the rescaled thickness of near-membrane 
layer $\delta'(t) = \delta(t)/ A$ with $A$ given by Eq.~(\ref{an})
for all values of $\kappa$, for glucose and for sucrose. The experimental
points are represented as in Fig.~2. According to Eqs.~(\ref{nml-sol},\ref{an}),
$\delta' (t)$ is simply the function $t^\gamma$, and, as seen in Fig.~9, all
our experimental data are indeed very well described by $t^{0.45}$.


\section{Final remarks}
\label{Final}


Our method to determine the parameters of subdiffusion relies on the 
near-membrane layers. One may ask why $\alpha$ and $D_{\alpha}$ are not 
extracted directly form the concentration profiles which are measured. 
There are three reasons to choose the near-membrane layers: experimental, 
theoretical and practical.

\begin{enumerate}

\item Measurement of the near-membrane-layer thickness does not suffer 
from the sizable ($\sim$ 10-15\%) systematic uncertainty of absolute 
normalization of the concentration profiles, as only the relative
concentration matters for $\delta$, see the definition (\ref{nml}).

\item Computed concentration profiles depend on the adopted boundary 
condition at a membrane while the condition is not well established 
even for the normal diffusion. The near-membrane layer is argued to be
free of this dependence.

\item When the concentration profile is fitted by a solution of the
subdiffusion equation, there are three free parameters: $\alpha$, 
$D_{\alpha}$ and the parameter characterizing the membrane permeability.
Because these fit parameters are correlated with each other it is very 
difficult in practice to get their unique values. One should remember 
here that the solution of the subdiffusion equation is of rather 
complicated structure, see Eqs.~(\ref{cpn}, \ref{cr}), which makes
the fitting procedure very tedious. When the temporal evolution of $\delta$ 
is discussed the membrane parameter drops out entirely, $\alpha$ is
controlled by the time dependence of $\delta (t)$ while $D_{\alpha}$ 
is provided by the coefficient $A$. 

\end{enumerate}

We have argued using Eq.~(\ref{se}) that $\delta$ scales as 
$t^\gamma$. A similar scaling can be obtained in diffusion equation 
with fractional derivatives in space and time if the orders of the
fractional operators are properly chosen. This is related to 
the fact that in fractional diffusion equation of order $\mu$
in space and order $\nu$ in time, $\langle x^2 \rangle$ scales
as $t^{2\nu/\mu}$. One could also come up with a nonlinear 
diffusion model without fractional derivatives that exhibits
the observed non-Gaussian scaling for $\delta$ \cite{non-lin}. 
This non-uniqueness problem cannot be solved in full generality but 
some possibilities can be excluded. The fractional differential 
equation of the order $\mu < 2$ of spatial derivative corresponds 
to `long' jumps (the variance is infinite) of the random walker. 
Since we study experimentally the diffusion in a gel built of large 
and heavy molecules forming a polymer network, long jumps are 
physically not plausible. We rather expect that the average waiting 
time of the random walker in a gel medium goes to infinity. This 
corresponds to the fractional differential equation with 
the temporal derivative of order $\alpha < 1$ as in Eq.~(\ref{se}). 
One cannot exclude the equation with fractional spatial {\em and} 
temporal derivatives. Similarly, the nonlinear diffusion equation 
cannot be excluded. However, Eq.~(\ref{se}), which is derived in 
a physically well motivated Continuous Time Random Walk approach 
\cite{mk,compte}, offers the simplest possibility. We also note that
the scaling shown in Fig.~9 strongly supports the subdiffusive
model based on Eq.~(\ref{se}). It seems difficult to reproduce
this scaling in a quite different theory. 
 
In our calculations presented in Sec~\ref{Theo}, the membrane is assumed
to be infinitely thin. Relaxing this assumption considerably complicates
theoretical analysis of the problem. This is not only the finite membrane
thickness that should be taken into account but the membrane internal 
structure should be modeled as well. In particular, one should answer 
the question whether transported substance is accumulated inside the 
membrane. If so, the membrane permeability might be time dependent. To 
avoid all these complications, the membrane is infinitely thin in our 
analysis which is a reasonable assumption as long as the membrane is 
sufficiently thin. (The thickness of the membrane used in our measurements 
was below 0.1 mm.) However, we expect that our method to determine the 
parameters $\alpha$ and $D_\alpha$ also works for finite width membranes. 
First of all, the temporal evolution of near-membrane layer was shown to 
be fully independent of the boundary condition at a membrane in the long 
time approximation. Our numerical calculations also suggest that $\delta (t)$ 
does not depend on the membrane. Finally, we observe that for the 
near-membrane-layer (\ref{nml}) only the relative substance concentration 
at $x=\delta$ and at the membrane surface ($x=0$) matters. Therefore,
it is expected that the membrane properties do not influence the time
evolution of near-membrane layer.

Our method to determine the subdiffusion parameters uses the membrane
system. While the membrane plays here only an auxiliary role, it should
be stressed that the transport in membrane systems is of interest in 
several fields of technology \cite{Rau89}, where the membranes are 
used as filters, and biophysics \cite{Tho76}, where the membrane transport 
plays a crucial role in the cell physiology. The diffusion in a membrane 
system is also interesting by itself as a nontrivial stochastic problem,
see {\it e.g.} \cite{kpr,kjp,kp1,kp}. Thus, our study of the subdiffusion 
in a membrane system, which to our best knowledge has not been investigated 
by other authors, opens up a new field of interdisciplinary research. 
It is also worth mentioning that our interferometric set-up can be used 
to experimentally study the (anomalous) diffusion not only in the membrane 
systems. In particular, we plan to perform measurements in a system with
no membrane where the sugar is transported directly from the water to gel 
solvent. However, there are problems to keep fixed geometry of the whole 
system in the course of long lasting diffusion process.

At the end let us summarize our considerations. We have developed a method
to extract the subdiffusion parameters $\alpha$ and $D_\alpha$ from 
experimental data. The method uses the membrane system, where the transported
substance diffuses from one vessel to another, and it relies on the fully 
analytic solution of the fractional subdiffusion equation. We have applied 
the method to our experimental data on the glucose and sucrose subdiffusion 
in a gel solvent and we have precisely determined the parameter $\alpha$ and
the subdiffusion coefficient $D_\alpha$.  

\acknowledgements

We are grateful to S\l awek W\c asik for help in performing the
measurements and in the data analysis.

\appendix


\section{Wall relation}
\label{appendix1}


We derive here the integral relation (\ref{gint}) which after the Laplace 
transformation reads
\begin{equation}
\label{gint-L}
\hat G_{+-}(x,s;x_0) = \hat J_{+-}(0^+,s;x_0) \:
\hat G_{\rm ref}(x,s;0^+) \;.
\end{equation}
We have introduced here the indices $+$ and $-$ which correspond to
the signs of $x$ and $x_0$.  Since $x > 0$ and $x_0 <0$, the Green's 
function is labeled with $+-$.

Because the current $J(x,s;x_0)$ is expressed, in accordance with 
Eq.~(\ref{current}), as
\begin{equation}
\hat J(x,s;x_0) = -D_{\alpha}\: s^{1-\alpha} \:
\frac{d\hat G(x,s;x_0)}{dx}\;,
\end{equation}
and the general solution of the subdiffusion equation is, as discussed
in the Appendix C, given by Eq.~(\ref{ltsfg}), the current equals
\begin{equation}
\label{cur77}
\hat J_{+-}(0^+,s;x_0) = \sqrt{D_\alpha} \, s^{1-\alpha/2} B(s) + 
\frac{1}{2} \, e^{x_0 \sqrt{\frac{s^\alpha}{D_\alpha}}}
= \sqrt{D_\alpha}\, s^{1-\alpha/2}\: \hat G_{+-}(0^+,s;x_0) \;.
\end{equation}
As the Green's function of the system with the reflecting wall 
at $x=0$ is given by Eqs.~(\ref{gref},\ref{go},\ref{golt}), 
we have
\begin{equation}
\hat G_{\rm ref}(x,s;0^+) = \frac{1}{\sqrt{D_{\alpha}}\, s^{1-\alpha/2}}
\: e^{-x \sqrt{\frac{s^{\alpha}}{D_{\alpha}}}}\;,
\end{equation}
and due to Eq.~(\ref{cur77}) the expression in the r.h.s. of 
Eq.~(\ref{gint-L}) equals
\begin{equation}
\label{78}
\hat J_{+-}(0^+,s;x_0) \:  \hat G_{\rm ref}(x,s;0^+) 
=  e^{-x \sqrt{\frac{s^{\alpha}}{D_{\alpha}}}}
\: \hat G_{+-}(0^+,s;x_0) \;.
\end{equation}

Eq.~(\ref{ltsfg}) allows one to write down the Green's function 
from the l.h.s. of Eq.~(\ref{gint-L}) as
\begin{equation}
\label{79}
\hat G_{+-}(x,s;x_0)  =  e^{-x \sqrt{\frac{s^{\alpha}}{D_{\alpha}}}}
\:  \hat G_{+-}(0^+,s;x_0) \;.
\end{equation}
Comparing Eq.~(\ref{78}) to  Eq.~(\ref{79}), one finally finds 
the wall relation (\ref{gint-L}).


\section{Function $\hat W$}
\label{appendix3}


We derive here Eq.~(\ref{W}). Taking the Laplace transform of the
boundary condition \ref{general-boundary}, replacing the concentration
profiles by the Green's functions and putting $x_0 <0$, we get
\be 
\label{general-boundary2}
b_1 \hat G_{--}(0^-,s;x_0) + b_2  \hat G_{+-}(0^-,s;x_0)  
+ b_3  \hat J(0,s;x_0)  = 0 \;.
\ee
Using Eq.~(\ref{ltsfg}), one finds 
\begin{eqnarray}
\label{g11}
\hat{G}_{--}(x,s;x_0) & = & A(s) \:
e^{x\sqrt{\frac{s^\alpha}{D_\alpha}}}
+ \frac{1}{2\sqrt{D_\alpha}\: s^{1-\frac{\alpha}{2}}} \:
e^{-\mid x-x_0\mid \sqrt{\frac{s^\alpha}{D_\alpha}}}\;,
\\ [2mm]
\label{g22}
\hat{G}_{+-}(x,s;x_0) & = & B(s) \:
e^{-x\sqrt{\frac{s^\alpha}{D_\alpha}}}
+ \frac{1}{2\sqrt{D_\alpha}s^{1-\frac{\alpha}{2}}} \:
e^{-\mid x-x_0\mid \sqrt{\frac{s^\alpha}{D_\alpha}}}\;.
\end{eqnarray}
The Green's functions (\ref{g11},\ref{g22}) provide, via 
Eq.~(\ref{current}), the currents
\begin{eqnarray}
\label{J11}
\hat{J}_{--}(0^-,s;x_0) & = & 
\frac{1}{2}\: e^{x_0\sqrt{\frac{s^\alpha}{D_\alpha}}}
- \sqrt{D_\alpha}\: A(s) \;,
\\ [2mm]
\label{J22}
\hat{J}_{+-}(0^+,s;x_0) & = & 
\frac{1}{2}\: e^{x_0\sqrt{\frac{s^\alpha}{D_\alpha}}}
+ \sqrt{D_\alpha}\: B(s) \;.
\end{eqnarray}
Due to the current continuity equation (\ref{con}), $A(s) = - B(s)$ and 
the current, which enters the boundary condition (\ref{general-boundary2}),
equals $\hat{J}_{+-}(0^+,s;x_0)$. Substituting the expressions 
(\ref{g11}, \ref{g22}, \ref{J22}) into Eq.~(\ref{general-boundary2}) and 
using the equality  $A(s) = - B(s)$, one finds
\be
B(s) = \frac{b_1 + b_2 + b_3  \sqrt{D_\alpha}\:s^{1-\alpha/2}}
{2\sqrt{D_\alpha}\:s^{1-\alpha/2} (b_1 - b_2 - b_3 \sqrt{D_\alpha}\:
s^{1-\alpha/2})} \: e^{x_0\sqrt{\frac{s^\alpha}{D_\alpha}}} \;.
\ee
The function $B(s)$ determines the current (\ref{J22}) which, 
substituted into the definition (\ref{W-def}) together with the
initial condition (\ref{ic}), provides the function $\hat W(s)$,
as given by Eq.~(\ref{W}).


\section{Solving subdiffusion equation}
\label{appendix2}


We briefly present here a procedure of solving the subdiffusion 
equation (\ref{se}) with the boundary conditions (\ref{con}, \ref{nbc}) 
and, respectively, (\ref{con}, \ref{rbc}). Taking the Laplace transform 
of Eq.~(\ref{se}), we obtain
\begin{equation}
\label{ltse}
\hat{C}(x,s)-s^{-\alpha}D_\alpha
\frac{d^2\hat{C}(x,s)}{dx^2}-C(x,0)=0\;,
\end{equation}
where, as previously, we use the hats to denote the Laplace transformed 
functions. The solution of Eq.~(\ref{ltse}) with the initial condition 
(\ref{icgf}) reads 
\begin{eqnarray}
\label{ltsfg}
\hat{G}( x,s;x_0)  =  A(s) \:  
e^{x\sqrt{\frac{s^\alpha}{D_\alpha}}} 
+ B(s) \: e^{-x\sqrt{\frac{s^\alpha}{D_\alpha}}} 
+  \frac{1}{2\sqrt{D_{\alpha}} \: s^{1-\frac{\alpha}{2}}}
e^{-\mid x-x_{0}\mid \sqrt{\frac{s^{\alpha}}{D_{\alpha}}}}\;,
\end{eqnarray}
where the functions $A(s)$ and $B(s)$ are determined by the boundary 
conditions. 


\subsection{Boundary conditions A}


The Laplace transform of boundary condition (\ref{nbc}) is
\begin{equation}
\label{ltnbc}
\hat{C}(0^{+},s)=\frac{1-\sigma}{1+\sigma} \: \hat{C}(0^{-},s)\,.
\end{equation}
Proceeding analogously to the case of normal diffusion \cite{kjp,kp1}, 
we find the Green's functions obeying the boundary condition (\ref{ltnbc}) 
as
\begin{eqnarray} 
\nonumber
\hat{G}_{--}( x,s;x_0)&=& \hat{G}_0(x,s;x_0)
+\sigma \hat{G}_0(-x,s;x_0)\;, 
\\ [2mm] \label{ngpp}
\hat{G}_{+-}(x,s;x_{0})&=&(1-\sigma) \: \hat{G}_0(x,s;x_0),
\end{eqnarray}
where $\hat{G}_0$ is the Laplace transform of Green's function 
for the homogeneous system without a membrane (\ref{go}); the indices 
$+$ and $-$ of the Green's functions refer to the sign of $x$ and of
$x_{0}$, respectively. 

To compute the concentration profiles we use the Laplace transformed 
of the integral relation (\ref{int}) which takes the form
\begin{equation}
\label{ltint}
\hat{C}(x,s) =\int \hat{G}(x,s;x_0) \: C(x_{0},0) dx_0\;.
\end{equation}
Now, we substitute the Green's function (\ref{ngpp}) into Eq.~(\ref{ltint}) 
and use the formula
\begin{equation}
\label{ge}
L^{-1}(s^\nu e^{-as^\beta}) =
\frac{1}{\beta a^{\frac{1+\nu}{\beta}}} \:
H_{1\;1}^{1\;0}\left( \frac{a^{\frac{1}{\beta}}}{t} \bigg|
\begin{array}{cc}
1 & 1 \\
\frac{1+\nu}{\beta} & \frac{1}{\beta}
\end{array}
\right) \;,
\end{equation}
which is derived in \cite{koszt04} using the Mellin transform technique 
\cite{s}. Here, $a,\beta >0$ while the parameter $\nu$ is not limited. After
simple calculations, we finally find the solution (\ref{cpn}).


\subsection{Boundary conditions B}


The subdiffusion equation (\ref{se}) is solved by the Green's functions 
of the form
\begin{eqnarray}
\label{g1}
\hat{G}_{--}(x,s;x_0) & = & A_1(s) \: 
e^{x\sqrt{\frac{s^\alpha}{D_\alpha}}} 
+ B_1(s) \: e^{-x\sqrt{\frac{s^\alpha}{D_\alpha}}} 
+ \frac{1}{2\sqrt{D_\alpha}\: s^{1-\frac{\alpha}{2}}} \:
e^{-\mid x-x_0\mid \sqrt{\frac{s^\alpha}{D_\alpha}}}\;,
\\ [2mm]
\label{g2}
\hat{G}_{+-}(x,s;x_0) & = & A_2(s) \: 
e^{x\sqrt{\frac{s^\alpha}{D_\alpha}}} 
+ B_2(s) \: e^{-x\sqrt{\frac{s^\alpha}{D_\alpha}}} 
+ \frac{1}{2\sqrt{D_\alpha}s^{1-\frac{\alpha}{2}}} \:
e^{-\mid x-x_0\mid \sqrt{\frac{s^\alpha}{D_\alpha}}}\;.
\end{eqnarray}
The functions $A_1(s)$, $A_2(s)$, $B_1(s)$, and $B_2(s)$ are now determined
by the boundary conditions (\ref{con},\ref{rbc}) which after the Laplace 
transformation take the form
\begin{eqnarray}
\label{ltcon}
\hat{J}(0^-,s;x_0) &=&\hat{J}(0^+,s;x_0)\;, \\[2mm]
\label{ltrbc}
\hat{J}(0,s;x_0) &=&\lambda \left(\hat{G}_{--}(0^-,s;x_0) 
-\hat{G}_{+-}(0^+,s;x_0)\right)\;,
\end{eqnarray}
with the Laplace transform of subdiffusive flux given 
by the formula \cite{z}
\begin{displaymath}
\hat{J}(x,s;x_0)=-D_{\alpha}\: s^{1-\alpha} \: 
\frac{d\hat{G}(x,s;x_0)}{dx}\;.
\end{displaymath}

For the infinite system, one demands vanishing of the Green's functions 
for $x\rightarrow \pm \infty$, which gives $B_1 \equiv A_2 \equiv 0$. 
Substituting the Green's functions (\ref{g1}, \ref{g2}) into the 
boundary conditions (\ref{ltcon}, \ref{ltrbc}), we obtain 
\begin{eqnarray}
\nonumber
\hat{G}_{--}(x,s;x_0)&=&
\frac{1}{4\lambda +2\sqrt{D_\alpha} \: s^{1-\frac{\alpha}{2}}} 
\: e^{\left( x+x_0\right) \sqrt{\frac{s^\alpha}{D_\alpha}}}
+\frac{1}{2\sqrt{D_\alpha} \: s^{1-\frac{\alpha}{2}}} \: 
e^{-\mid x-x_{0}\mid \sqrt{\frac{s^\alpha}{D_\alpha}}}\;,
\\[3mm]\label{r1}
\hat{G}_{+-}(x,s;x_0)&=&
\frac{4\lambda}{4\lambda +2\sqrt{D_\alpha} \: s^{1-\frac{\alpha}{2}}}
\times \frac{1}{2\sqrt{D_\alpha} \: s^{1-\frac{\alpha}{2}}} 
e^{-\left( x-x_0\right) \sqrt{\frac{s^\alpha}{D_\alpha}}}\;.
\end{eqnarray}
Expanding the function (\ref{r1}) into the power series with respect to $s$
and using the initial condition (\ref{ic}), the integral relation (\ref{ltint})
provides, with the help of the formula (\ref{ge}), the solution (\ref{cr}).



\newpage

\begin{figure}

\hspace{2cm}
\epsfig{file=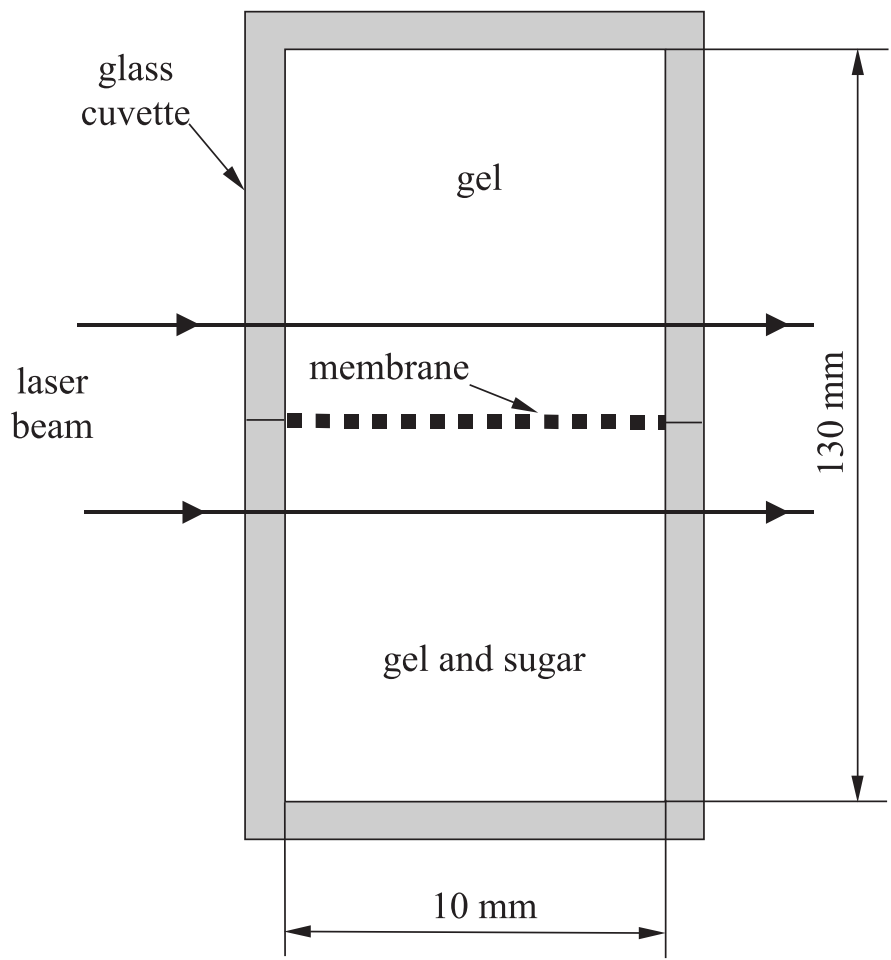,width=120mm}

\caption{Schematic view of the membrane system under study.}

\end{figure}


\begin{figure}

\hspace{2cm}
\epsfig{file=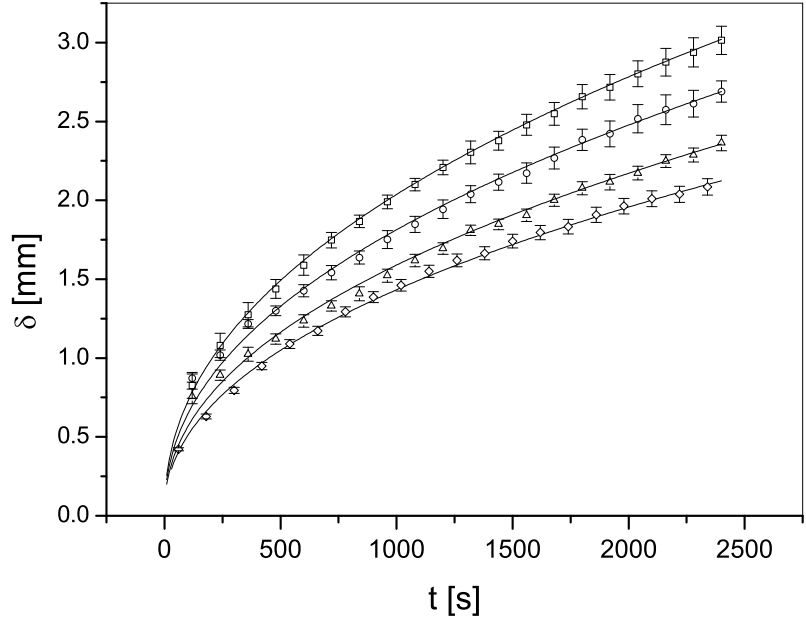,width=120mm}

\caption{The experimentally measured thickness of the near-membrane layer
$\delta$ as a function of time $t$ for glucose with $\kappa=0.05$ ($\Box$), 
$\kappa=0.08$ ($\circ$), $\kappa=0.12$ ($\triangle$), and for sucrose with 
$\kappa=0.08$ ($\diamondsuit$). The lines represent the power function 
$A \: t^{0.45}$ with the coefficients $A$ given by 
Eqs.~(\protect\ref{A-g1}-\protect\ref{A-s1}).}

\end{figure}


\begin{figure}

\hspace{2cm}
\epsfig{file=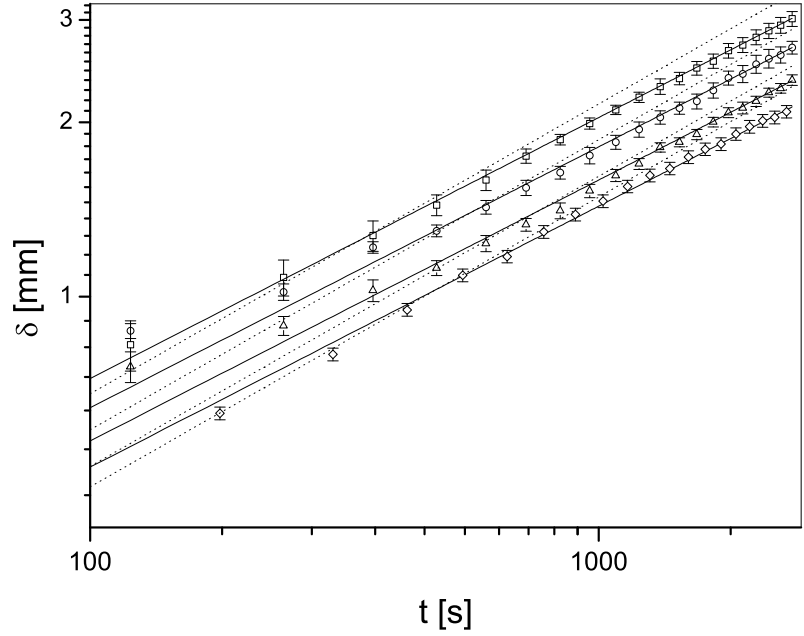,width=120mm}

\caption{The same experimental data as in Fig.~2 but in a log-log scale.
The solid lines represent the power function $A \: t^{0.45}$ with the 
coefficients $A$ given by Eqs.~(\protect\ref{A-g1}-\protect\ref{A-s1}). 
The dotted lines correspond to the function $A \sqrt{t}$.} 

\end{figure}


\begin{figure}

\hspace{2cm}
\epsfig{file=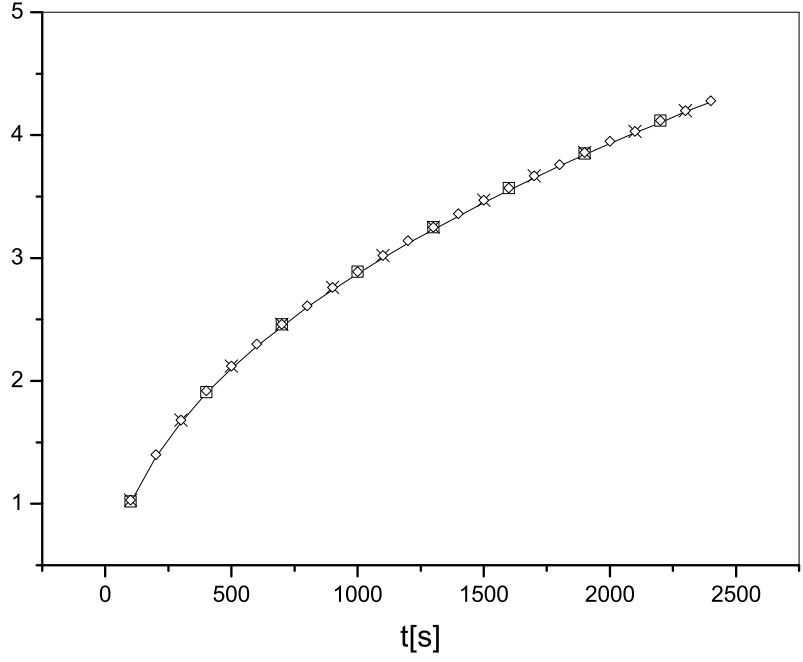,width=120mm}

\caption{The time evolution of the near-membrane layer $\delta_{\rm B}$ 
obtained from Eq.~(\protect\ref{cr}) for different values of the membrane 
permeability parameter $\lambda$: 1 [mm/s] (solid line), 5 [mm/s] ($\Box$), 
100 [mm/s] ($\times$), and 500 [mm/s] ($\diamondsuit$). The remaining 
parameters equal $\kappa=0.12$, $\alpha=0.90$, and 
$D_{0.90} = 3\times10^{-3}\;{\rm mm^2/s^{0.90}}$.} 

\end{figure}


\begin{figure}

\hspace{2cm}
\epsfig{file=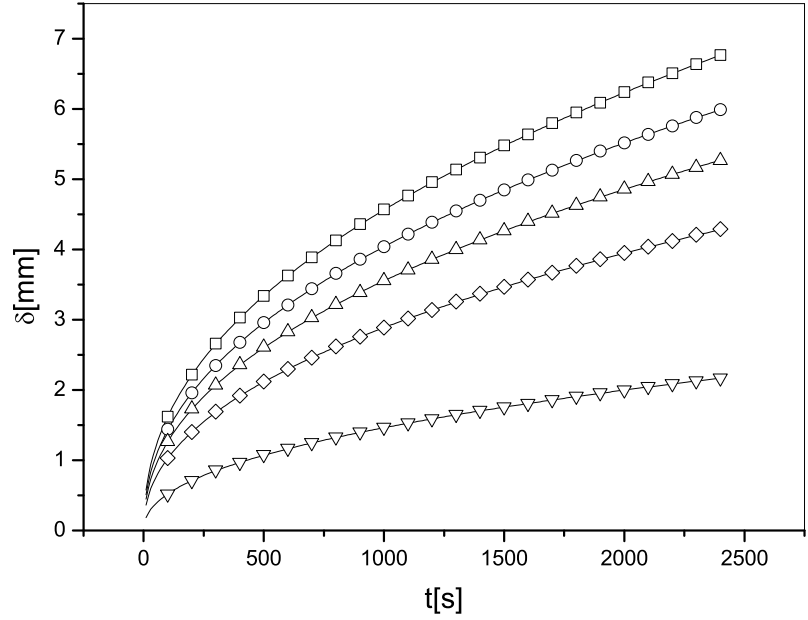,width=120mm}

\caption{Comparison of $\delta_{\rm A}(t)$ (symbols) and $\delta_{\rm B}(t)$ 
(lines) for several values of $\kappa$: 0.05 ($\Box$), 0.08 ($\circ$), 
0.12 ($\triangle$), 0.20 ($\diamondsuit$), 0.50 ($\bigtriangledown$). 
The remaining parameters equal $\alpha = 0.90$ and 
$D_{0.90} = 3\times10^{-3}\;{\rm mm^2/s^{0.90}}$.}

\end{figure}


\begin{figure}

\hspace{2cm}
\epsfig{file=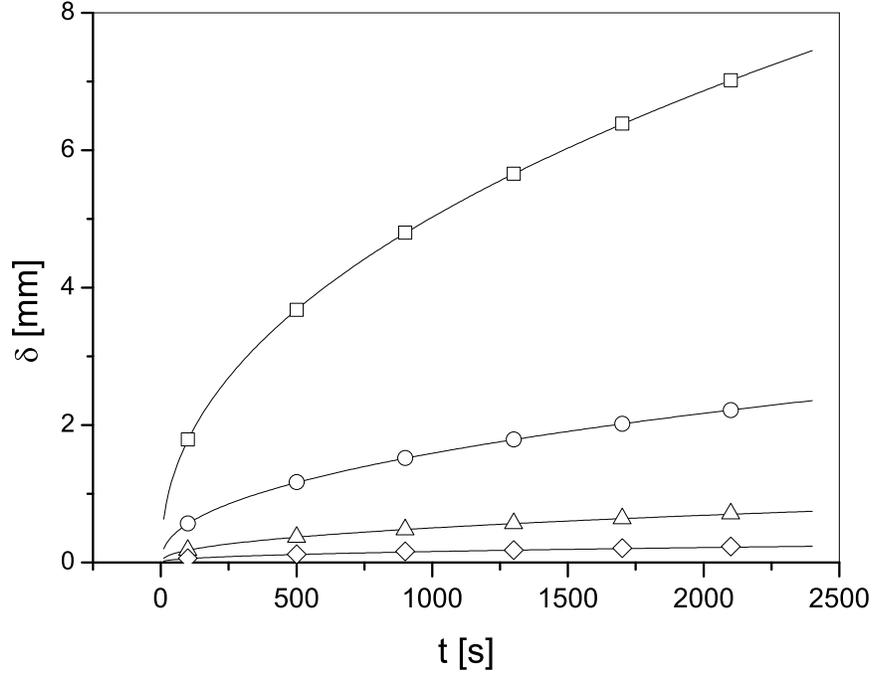,width=130mm}

\caption{Comparison of $\delta_{\rm A}(t)$ (symbols) and $\delta_{\rm B}(t)$
(lines) for several values of $D_{0.90}$: 
$1\times10^{-2}\;{\rm mm^2/s^{0.9}}$ ($\Box$), 
$1\times10^{-3}\;{\rm mm^2/s^{0.9}}$ ($\circ$), 
$1\times10^{-4}\;{\rm mm^2/s^{0.9}}$ ($\triangle$), 
$1\times10^{-5}\;{\rm mm^2/s^{0.9}}$ ($\diamondsuit$).
The remaining parameters equal $\alpha = 0.90$ and $\kappa = 0.12$.}

\end{figure}


\begin{figure}

\hspace{2cm}
\epsfig{file=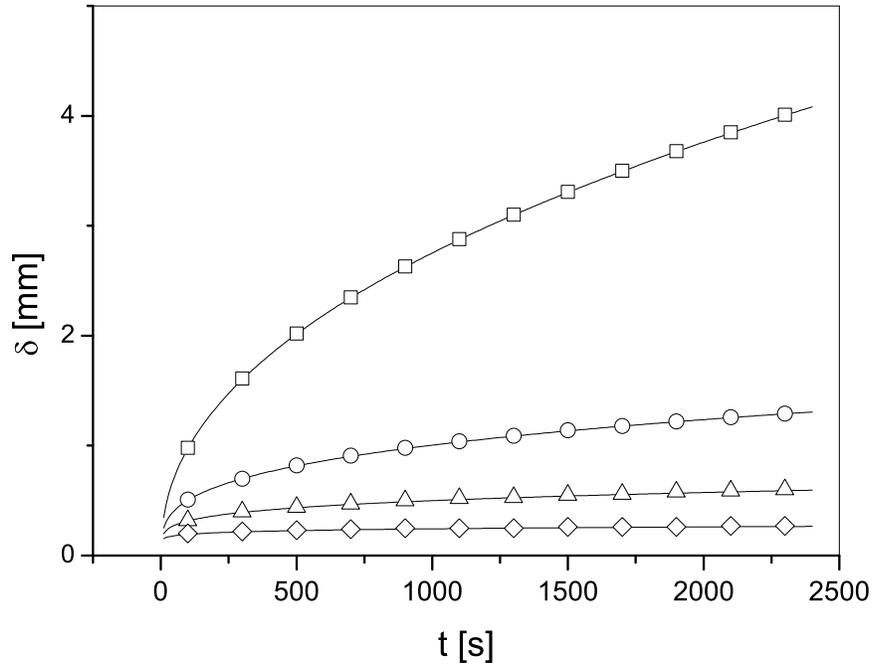,width=130mm}

\caption{Comparison of $\delta_{\rm A}(t)$ (symbols) and $\delta_{\rm B}(t)$
(lines) for several values of $\alpha$: 0.9 ($\Box$), 0.6 ($\circ$), 0.4 
($\triangle$), 0.2 ($\diamondsuit$). The remaining parameters equal 
$\kappa=0.12$ and
$D_\alpha = 3\times10^{-3}\;{\rm mm^2/s^\alpha}$.}

\end{figure}


\begin{figure}

\hspace{2cm}
\epsfig{file=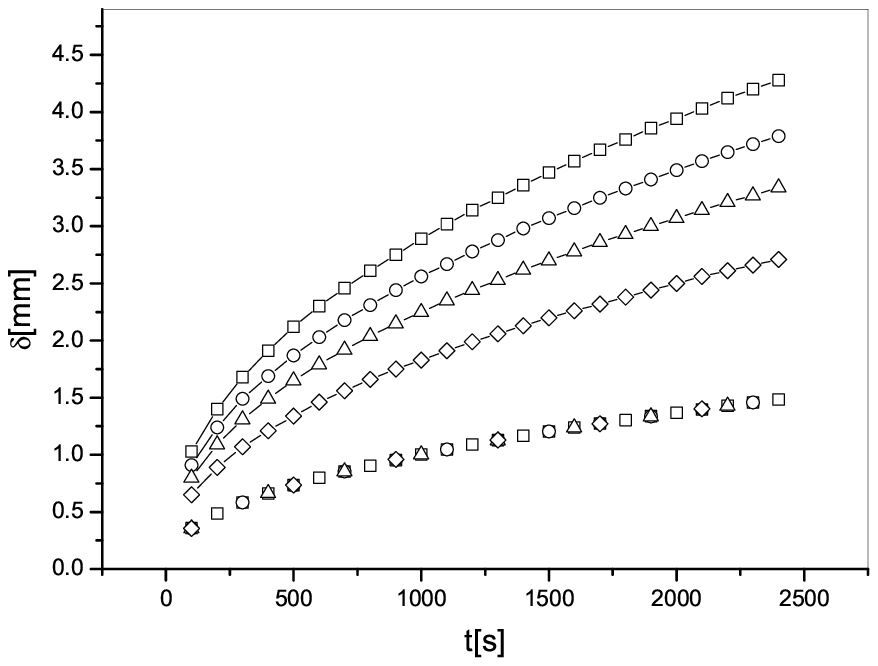,width=120mm}

\caption{The thickness of the near-membrane layer $\delta_{\rm B}$ 
and the rescaled thickness $\delta_{\rm B}/A_{\rm A}$ as functions of 
time for several values of $\kappa$: 0.05 ($\Box$), 0.08 ($\circ$), 
0.12 ($\triangle$), 0.20 ($\diamondsuit$). For the rescaled thickness 
the symbols are on top of each other. The parameters $\alpha$ and 
$D_\alpha$ equal, respectively, 0.90 and 
$3\times10^{-3}\;{\rm mm^2/s^\alpha}$.} 

\end{figure}


\begin{figure}

\hspace{2cm}
\epsfig{file=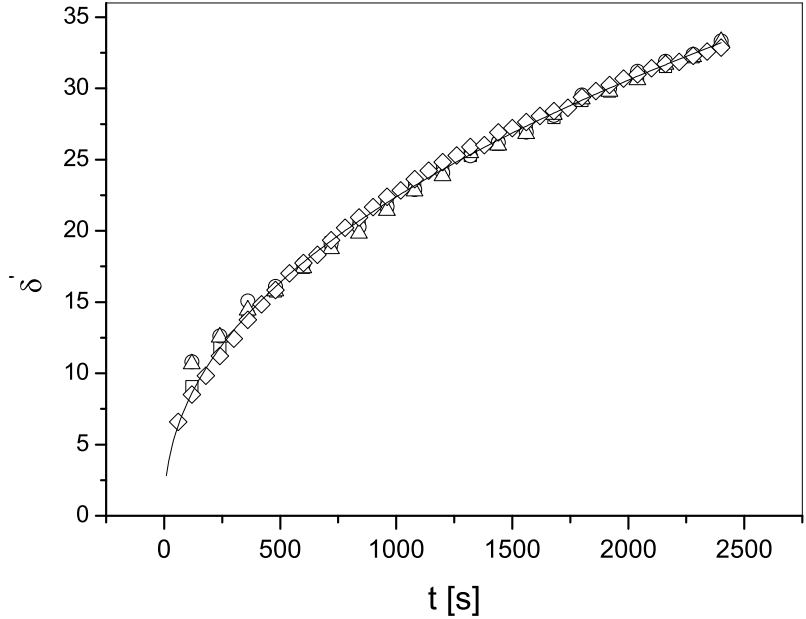,width=120mm}

\caption{The experimentally measured $\delta$ divided by the coefficient
$A$ given by Eq.~(\protect\ref{an}). The symbols are assigned as in
Fig.~2. The parameters equal $\alpha = 0.90$ and 
$D_{0.90}= 1.0 \times 10^{-3}\; [{\rm mm^2/s^{0.90}}]$ for glucose and
$D_{0.90}= 6.3 \times 10^{-3}\; [{\rm mm^2/s^{0.90}}]$ for sucrose.
For clarity of the plot the error bars are not shown. The line represents
the function $t^{0.45}$.} 

\end{figure}

\end{document}